\documentclass[a4paper,12pt]{article}

\usepackage{graphicx}
\usepackage{array}
\usepackage{amsmath}
\usepackage{amssymb}
\usepackage{latexsym}
\usepackage{subfigure}
\usepackage{color}

\makeatletter
\@addtoreset{equation}{section}
\makeatother

\date{empty}
\begin{document}
\begin{titlepage}
\null
\begin{flushright}
June, 2014
\end{flushright}
\vskip 0.5cm
\begin{center}
  {\Large \bf BPS States in Supersymmetric  Chiral Models \\
\vskip 0.3cm
with Higher Derivative Terms
}
\vskip 1.1cm
\normalsize
\renewcommand\thefootnote{\alph{footnote}}

{\large
Muneto Nitta$^{\dagger}$\footnote{nitta(at)phys-h.keio.ac.jp}
and Shin Sasaki$^\ddagger$\footnote{shin-s(at)kitasato-u.ac.jp}
}
\vskip 0.7cm
  {\it
  $^\dagger$ 
Department of Physics, and Research and Education Center for Natural Sciences, \\
\vskip -0.2cm
Keio University, Hiyoshi 4-1-1, Yokohama, Kanagawa 223-8521, Japan  
\vskip 0.1cm
$^\ddagger$
  Department of Physics,  Kitasato University \\
  \vskip -0.2cm
  Sagamihara 252-0373, Japan
}
\vskip 0.5cm
\begin{abstract}
We study the higher derivative chiral models with four supercharges 
and Bogomol'nyi-Prasad-Sommerfield (BPS) states in these models.
The off-shell Lagrangian generically includes higher powers of the
 auxiliary fields $F$ which causes distinct on-shell branches associated with the solutions to
 the auxiliary fields equation.
We point out that the model admits a supersymmetric completion of 
arbitrary higher derivative bosonic models of a
single complex scalar field and 
an arbitrary scalar potential can be 
introduced even without superpotentials. 
As an example,  we present 
a supersymmetric extension of the 
Faddeev-Skyrme model 
without four time derivatives, 
in contrast to the previously proposed supersymmetric  
Faddeev-Skyrme-like model 
containing four time derivatives. 
In general, higher derivative terms together with a superpotential 
result in deformed scalar potentials.
We find that higher derivative corrections to 1/2 BPS domain
 walls and 1/2 BPS lumps are exactly canceled out while the 1/4 BPS lumps (as compact baby Skyrmions) depend on a
 characteristic feature of the higher derivative models.
We also find a new 1/4 BPS condition for domain 
 wall junctions which generically receives higher derivative corrections.
\end{abstract}
\end{center}

\end{titlepage}

\newpage
\setcounter{footnote}{0}
\renewcommand\thefootnote{\arabic{footnote}}
\pagenumbering{arabic}
\section{Introduction}
Low energy dynamics of field theories 
can be described by only light fields such as 
Nambu-Goldstone modes 
when one integrates out 
massive modes. 
The low-energy effective theories are 
usually expanded by derivative expansions;
thereby, they inevitably contain higher derivative terms of fields. 
Chiral perturbation theory is such a theory 
describing low-energy pion dynamics in QCD with 
a chiral symmetry breaking \cite{Leutwyler:1993iq}. 
The Skyrme model \cite{Sk}, which is a non-linear sigma model with fourth
order derivative terms, 
is one of such a class. 
Supergravity as low-energy effective theory of string theory 
should have higher derivative correction terms \cite{Polchinski:1998rq}.
Other examples include world-volume effective actions of solitonic objects such as 
 topological solitons in field theories 
and D-branes in string theories \cite{FrTs}. 
The effective theory of a D-brane is described by 
the Dirac-Born-Infeld (DBI) action \cite{DiBoIn}
containing an infinite number of derivatives. 
Higher derivative field theories 
are also useful in other areas of physics. 
In the cosmological context, higher derivative theories are proposed for
inflation models such as the K-inflation \cite{ArDaMu} 
and the Galileon inflation \cite{NiRaTr}.
These higher derivative models are known to 
admit characteristic soliton solutions such
as k-defects \cite{Ba}, compactons \cite{Adam:2009px,AdRoSaGuWe} and so on.

On the other hand, supersymmetry is 
one of the most important tools in modern high energy physics. 
It has not only been considered  as the most promising candidate to solve the naturalness problem in the 
Standard Model
 in the phenomenological side, 
but also it plays important roles 
to control quantum corrections 
in supersymmetric field theories, leading  
to determining exact low-energy dynamics \cite{Seiberg:1994rs}.  
When one constructs low-energy effective theories 
in supersymmetric field theories, 
one is required to consider higher derivative corrections 
in a supersymmetric manner.
It is, however, not so easy to construct supersymmetric
completion of general higher derivative theories. 
Off-shell superfield formalisms are useful to write down actions of supersymmetric higher derivative models. 
In particular, the four-dimensional $\mathcal{N} = 1$ superfield formalism
that incorporates the chiral superfield $\Phi$ is a simple starting point.
It is, however, 
known that not all the off-shell supersymmetric higher derivative models exhibit good physical
properties.
Off-shell formulations of 
higher derivative terms often encounter with 
an auxiliary field problem;  
chiral superfields with space-time derivatives
(e.g. $\partial_m \Phi$) sometimes introduce derivative interactions of
the auxiliary field $F$.
Consequently, the auxiliary fields become dynamical.
It is hard to eliminate them, and 
 the on-shell structure of the action 
is not obvious.
For instance, the chiral Lagrangian of QCD contains 
the Wess-Zumino-Witten (WZW) term 
to reproduce the quantum anomaly at low energy. 
However, a supersymmetric completion 
of the WZW term  
proposed in Ref.~\cite{Nemeschansky:1984cd} 
suffers from this auxiliary field problem 
\cite{Gates:1995fx,Nitta:2001rh}.
It was proposed in Ref.~\cite{Gates:1998si} that 
a supersymmetric WZW term in superspace 
 can be constructed 
without the auxiliary field problem 
if the number of chiral superfields is doubled.\footnote{
The actual form of the WZW term was derived 
in Refs.~\cite{Gates:2000rp,Banin:2006db} 
and includes a K\"ahler tensor discussed in 
the next section.
}
The auxiliary field problem 
would be more problematic if one were to introduce a superpotential, 
so one could not introduce a potential.

Nevertheless, 
supersymmetric higher derivative models of which the building blocks are the chiral
superfields are studied in various contexts. 
Among other things, the chiral models studied in Refs.~\cite{AdQuSaGuWe, KhLeOv} 
provide a good grounding for studying supersymmetric
higher derivative theories. 
In this model, the auxiliary fields are not accompanied by the
space-time derivatives and therefore they can be 
eliminated by their equations
of motion. 
In principle, it is possible to write down the explicit on-shell
actions of the models. 
In particular, the scalar potential that shows
up after eliminating the auxiliary fields looks more apparent
\cite{SaYaYo}. 
The coupling of higher derivative chiral models to supergravity was also achieved in this type of model 
\cite{KoLeOv, FaKe}.
A supersymmetric DBI action was constructed in Ref.~\cite{RoTs}.
The other examples include 
a supersymmetric completion of the $P(X,\varphi)$ model \cite{KhLeOv}, 
the supersymmetric Galileon inflation models \cite{KhLeOv2} 
and models for the ghost condensation \cite{KoLeOv2}. 
The same structure appears in quantum effective actions \cite{Buchbinder:1994iw, Kuzenko:2014ypa}
A higher derivative supersymmetric 
${\mathbb C}P^1$ model 
free from the auxiliary field problem
was also considered previously 
as a supersymmetric extension 
\cite{BeNeSc,Fr} of the Faddeev-Skyrme model \cite{Faddeev:1996zj} and a
supersymmetric baby Skyrme model \cite{Adam:2011hj,AdQuSaGuWe}. 
The formalism in Refs.~\cite{AdQuSaGuWe, KhLeOv} has been also applied
to the construction of manifestly supersymmetric
higher derivative corrections to supersymmetric
nonlinear realizations \cite{Nitta:2014fca}.

In the former half of this paper, 
we study higher derivative chiral models 
developed in Refs.~\cite{AdQuSaGuWe, KhLeOv} 
in the superfield formalism, 
where higher derivative terms can be introduced 
as a tensor with two holomorphic and symmetric indices 
and two anti-holomorphic and symmetric indices.
We find a surprising fact that has  
been overlooked in past studies on the supersymmetric
completions of various higher derivative models.
The model with a single chiral superfield 
admits a supersymmetric extension of {\it arbitrary} bosonic models that consist of a single complex scalar field. 
As an example, we present 
a supersymmetric extension of the 
Faddeev-Skyrme model \cite{Faddeev:1996zj}. 
The bosonic part of this model does 
not contain four time derivatives.
This is in contrast to the
previously proposed supersymmetric extension 
\cite{BeNeSc,Fr} of the Faddeev-Skyrme model 
that contains an additional four derivative term 
that includes four time derivatives.
Moreover, we point out 
that an arbitrary scalar potential can be 
introduced even without the superpotential. 
We further work out the
higher derivative chiral models with superpotentials. 
The resulting on-shell Lagrangians are highly non-linear.
We study perturbative analysis revealing 
the possibility of ghost kinetic term and
deformations of the scalar potential.

Meanwhile, Bogomol'nyi-Prasad-Sommerfield (BPS) 
topological solitons play important roles 
in the study of non-perturbative dynamics of 
supersymmetric field theories 
since they break and preserve a fraction of supersymmetry, belong to short supermultiplets, 
and consequently are stable against 
quantum corrections \cite{Witten:1978mh}. 
When a BPS soliton preserves $p/q$ of supersymmetry, 
it is called a $p/q$ BPS soliton.
For instance, Yang-Mills instantons, 
BPS monopoles, vortices, lumps and 
domain walls \cite{Dvali:1996xe} are of $1/2$ BPS 
and composite solitons such as 
domain wall junctions are of 1/4 BPS 
in theories with four supercharges \cite{GiTo,Oda:1999az,NaNiSa}
and eight supercharges \cite{Eto:2005cp} 
(see Refs.~\cite{Shifman:2007ce,Eto:2006pg,Eto:2005sw} 
as a review for a fraction of supersymmetry for BPS states). 
BPS solitons remain important 
in supersymmetric 
field theories with higher derivative terms. 
Prime examples of such 
solitons contain 1/2 BPS lumps in supersymmetric 
${\mathbb C}P^1$ models with a 
four-derivative term \cite{Eto:2012qda}, 
supersymmetric baby Skyrmions, which are compactons  \cite{Adam:2011hj,AdQuSaGuWe};
and BPS compactons in K-field theories 
\cite{AdQuSaGuWe2,AdQuSaGuWe3}.
The higher derivative ${\mathbb C}P^1$ model 
in Ref.~\cite{Eto:2012qda}  
appears as the effective theory of  
a 1/2 BPS non-Abelian vortex 
\cite{Hanany:2003hp} 
in supersymmetric theories with eight supercharges. 
Then, the 1/2 BPS lumps in the vortex 
correspond to Yang-Mills instantons in the bulk 
\cite{Eto:2004rz}. 
While a few examples of BPS solitons in 
higher derivative supersymmetric theories 
have been studied thus far, 
a systematic study of BPS solitons 
in such theories is needed.

In the latter half of this paper, we give a general framework to 
examine BPS states in supersymmetric higher derivative
chiral models. Our framework does not only reproduce, 
in a unified manner, a few remarkable previous studies of 
the BPS bounds in the supersymmetric higher
derivative models admitting BPS baby Skyrmions \cite{Adam:2011hj,AdQuSaGuWe}, 
BPS compactons \cite{AdQuSaGuWe2,AdQuSaGuWe3},
and BPS lumps \cite{Eto:2012qda},  
but also includes the more general cases with several new BPS states; 
1/2 BPS domain walls, 
1/4 BPS domain wall junctions, 
1/2 and 1/4 BPS lumps and baby Skyrmions. 
In particular, we find 
BPS baby Skyrmions found in Ref.~\cite{AdQuSaGuWe} 
to be 1/4 BPS states. 
We show that 1/2 BPS domain walls 
and 1/2 BPS lumps do not receive 
higher derivative corrections 
while 1/4 BPS domain wall junctions do.

The organization of this paper is as follows.
In Sec.~\ref{sec:hdc}, we introduce the supersymmetric higher derivative
chiral model with four supercharges. 
We write down the equation of motion for the auxiliary fields
and analyze the structure of the on-shell Lagrangians. 
In particular, we introduce the superpotential and the deformation of
the scalar potential caused by the higher derivative terms is discussed.
 We then examine BPS states that preserve 1/2 and 1/4 of the
original supersymmetry in subsequent sections. 
The 1/2 BPS domain wall and 1/4 BPS 
domain wall junctions 
are studied in Sec.~\ref{sec:wall},
and 1/2 BPS and 1/4 BPS lumps are studied in Sec.~\ref{sec:lump}. 
Section \ref{sec:conc} is devoted to conclusion and
discussions. Notations and conventions of superfields are found in
the Appendix \ref{sec:notation}.

\section{Higher derivative chiral models}\label{sec:hdc}
In the first subsection, we present 
general higher derivative chiral models 
with multiple chiral superfields. 
In the second subsection,
we further work out the models with 
a single chiral superfield without and with 
a superpotential.
\subsection{General chiral models}
We consider four-dimensional $\mathcal{N} = 1$
supersymmetric higher derivative chiral models that have specific
properties. 
The Lagrangian consists of chiral superfields $\Phi^i$ $(i=1, \cdots,
N)$, for which the component expansion in the chiral base $y^m =
x^m + i \theta \sigma^m \bar{\theta}$ is 
\begin{align}
\Phi^i (y,\theta) = \varphi^i (y) 
+ \theta \psi^i (y) + \theta^2 F^i(y),
\end{align}
where $\varphi^i$ is the complex scalar field, $\psi^i$ is the Weyl
fermion and $F^i$ is the complex auxiliary field.
The notations and conventions of the chiral superfield are found in
Appendix 
\ref{sec:notation}.

The supersymmetric Lagrangian with 
higher derivative terms is given by 
\begin{align}
\mathcal{L} =& \ \int \! d^4 \theta \ K (\Phi^i, \Phi^{\dagger \bar{j}}) 
+ \frac{1}{16} \int \! d^4 \theta \ 
\Lambda_{ik\bar{j} \bar{l}}
(\Phi,
 \Phi^{\dagger}) 
D^{\alpha} \Phi^i
 D_{\alpha} \Phi^k \bar{D}_{\dot{\alpha}} \Phi^{\dagger \bar j}
 \bar{D}^{\dot{\alpha}} \Phi^{\dagger \bar{l}} 
\notag \\
& + \left(\int \! d^2 \theta \ W(\Phi^i) + {\rm h.c.}\right)
\label{eq:Lagrangian}
\end{align}
where $K$ is the K\"ahler potential 
and 
$W$ is a superpotential as usual.
Higher derivative terms are produced by 
the second term proportional to  
$\Lambda_{ik\bar{j} \bar{l}}$, 
which is 
a $(2,2)$ K\"ahler tensor 
symmetric in holomorphic and anti-holomorphic indices, 
of which the components are 
functions of $\Phi^i$ and $\Phi^{\dagger \bar{i}}$
(admitting space-time derivatives acting on them).\footnote{
This tensor term was obtained 
in Ref.~\cite{Banin:2006db} 
as a part of the supersymmetric Wess-Zumino-Witten term. 
}
As we will see, the most important 
feature  
of this model is that the auxiliary fields never become
dynamical; the equation of motion for the auxiliary fields is an algebraic equation.

Now we examine the component structure of the model \eqref{eq:Lagrangian}.
The fourth derivative part of the Lagrangian \eqref{eq:Lagrangian} has 
an essential property. This term is evaluated as 
\begin{align}
D^{\alpha} \Phi^i D_{\alpha} \Phi^k \bar{D}_{\dot{\alpha}}
 \Phi^{\dagger\bar{j}} \bar{D}^{\dot{\alpha}}
 \Phi^{\dagger\bar{l}} 
=& \ 16 \theta^2 \bar{\theta}^2 
\left[
\frac{}{}
(\partial_m \varphi^i \partial^m \varphi^k) (\partial_m
 \bar{\varphi}^{\bar{j}} \partial^m \bar{\varphi}^{\bar{l}})
- 2
\partial_m \varphi^i F^k 
\partial^n \bar{\varphi}^{\bar{j}} \bar{F}^{\bar{l}} 
+ F^i \bar{F}^{\bar{j}} F^k \bar{F}^{\bar{l}}
\right]
+ I_f, 
\label{eq:4th}
\end{align}
where $I_f$ stands for terms that contain
 fermion fields.
Since the bosonic part 
of the right hand side of \eqref{eq:4th} saturates
the Grassmann coordinate $\theta^2 \bar{\theta}^2$, 
only the lowest component of the tensor $\Lambda_{ik\bar{j} \bar{l}}$
contributes to the bosonic part of the Lagrangian.
Therefore the bosonic part of the Lagrangian \eqref{eq:Lagrangian} is 
\begin{align}
\mathcal{L}_b =& \ 
\frac{\partial^2 K}{\partial \varphi^i \partial \bar{\varphi}^{\bar{j}}} 
(- \partial_m \varphi^i \partial^m \bar{\varphi}^{\bar{j}} + F^i
 \bar{F}^{\bar{j}} )
+ \frac{\partial W}{\partial \varphi^i} F^i + \frac{\partial
 \bar{W}}{\partial \bar{\varphi}^{\bar{j}}} \bar{F}^{\bar{j}}
\notag \\
& + \Lambda_{ik\bar{j} \bar{l}} (\varphi, \bar{\varphi})
\left[
\frac{}{}
(\partial_m \varphi^i \partial^m \varphi^k) (\partial_n
 \bar{\varphi}^{\bar{j}} \partial^n \bar{\varphi}^{\bar{l}})
-
\partial_m \varphi^i F^k 
\partial^m \bar{\varphi}^{\bar{j}} \bar{F}^{\bar{l}} 
+ F^i \bar{F}^{\bar{j}} F^k \bar{F}^{\bar{l}}
\right].
\label{eq:comLagrangian}
\end{align}
This Lagrangian exhibits a higher derivative model that has the following properties: 
(I) the higher derivative terms are governed by the tensor
$\Lambda_{ik\bar{j} \bar{l}}$, and (II) the model is manifestly
(off-shell) supersymmetric and K\"ahler invariant provided that $K$ and $W$
are scalars and $\Lambda_{ik\bar{j} \bar{l}}$ is a tensor.
Among other things, the auxiliary fields do not have a space-time
derivative\footnote{This is true only for the purely bosonic terms. 
There are derivative interactions of the auxiliary fields in the
fermionic contributions $I_f$ \cite{KhLeOv}. 
They are irrelevant when classical configurations of fields are concerned.
} 
and they are eliminated by the following equation of motion:
\begin{align}
\frac{\partial^2 K}{\partial \varphi^i \partial \bar{\varphi}^{\bar{j}}} 
F^i - 2 \partial_m \varphi^i F^k \Lambda_{ik\bar{j} \bar{l}} 
 \partial^m \bar{\varphi}^{\bar{l}} + 
2 \Lambda_{ik\bar{j} \bar{l}} F^i F^k \bar{F}^{\bar{l}} 
+ \frac{\partial \bar{W}}{\partial \bar{\varphi}^{\bar{j}}} 
= 0. \label{eq:af-eom}
\end{align}
This is an algebraic equation and, in principle, solvable.
However, the equation \eqref{eq:af-eom} is a simultaneous equation of
cubic power and it is hard to find explicit solutions $F_i$.
We comment that when $W=0$ at least $F_i = 0$ is a solution.
In this case, the on-shell Lagrangian becomes 
\begin{align}
\mathcal{L}_b = - \frac{\partial^2 K}{\partial \varphi^i \partial
 \bar{\varphi}^{\bar{j}}} \partial_m \varphi^i \partial^m
 \bar{\varphi}^{\bar{j}} 
+ \Lambda_{i k \bar{j} \bar{l}} 
(\partial_m \varphi^i \partial^m \varphi^k) (\partial_n \bar{\varphi}^{\bar{j}}
 \partial^n \bar{\varphi}^{\bar{l}}).
\end{align}
In general, there are more solutions other than $F_i = 0$ 
which we will show explicitly for models with one component field.

\subsection{Chiral models of one component}
Now we consider the single chiral superfield $\Phi$ for simplicity. 
The equation of motion for the auxiliary field becomes
\begin{eqnarray}
\begin{aligned}
 & K_{\varphi \bar{\varphi}} F 
- 2 F 
\left(
 \partial_m \varphi \partial^m \bar{\varphi} - F \bar{F}
\right) \Lambda (\varphi, \bar{\varphi}) 
+ \frac{\partial \bar{W}}{\partial
\bar{\varphi}} = 0.  \label{eq:af-eom2}
\end{aligned}
\label{eq:aux_eq}
\end{eqnarray}
Here $K_{\varphi \bar{\varphi}} = \frac{\partial K}{\partial \varphi
\partial \bar{\varphi}}$.
We solve the equation \eqref{eq:aux_eq} in the $W=0$ and $W\not=0$ cases separately.

\subsubsection{$W = 0$ case}
When there is no superpotential, the equation for the auxiliary field becomes
\begin{align}
K_{\varphi \bar{\varphi}} F
- 2 F 
\left(
\partial_m \varphi \partial^m \bar{\varphi} - F \bar{F}
\right) \Lambda 
= 0.
\label{eq:eom_auxiliary_no_supot}
\end{align}
Then the solutions are found to be 
\begin{align}
F
=& \ 0, 
\label{eq:first_sol}
\\
F \bar{F} =& \ - \frac{K_{\varphi \bar{\varphi}}}{2 \Lambda} + \partial_m \varphi \partial^m
 \bar{\varphi}.
\label{eq:second_sol}
\end{align}
There are 
two different on-shell branches associated with
the solutions \eqref{eq:first_sol} and \eqref{eq:second_sol}.

For the first solution \eqref{eq:first_sol}, the bosonic part of the on-shell Lagrangian is 
\begin{align}
\mathcal{L}_{1b} = - K_{\varphi \bar{\varphi}} \partial_m \varphi \partial^m \bar{\varphi} + 
(\partial_m \varphi \partial^m \varphi) (\partial_n \bar{\varphi}
 \partial^n \bar{\varphi}) \Lambda.
\label{eq:W0_osl1}
\end{align}
The first term is the ordinary kinetic term and the second term contains
higher derivative correction terms. We call this the canonical branch.

An example of the model is the $\mathcal{N} = 1$ supersymmetric DBI
action for the world-volume theory of single D3-brane.
The corresponding K\"ahler metric is canonical, $K_{\varphi \bar{\varphi}} = 1$, and
the function $\Lambda$ 
is given by \cite{RoTs}
\begin{align}
\Lambda = 
\frac{1}{
1 + A + \sqrt{(1 + A)^2 - B}
}, \quad 
A = \partial_m \Phi \partial^m \Phi^{\dagger}, \quad 
B = \partial_m \Phi \partial^m \Phi \partial_n \Phi^{\dagger} \partial^n \Phi^{\dagger}.
\end{align}
The other examples include 
a supersymmetric completion of the $P(X,\varphi)$ model \cite{KhLeOv}, 
the supersymmetric Galileon inflation models \cite{KhLeOv2} 
and models for the ghost condensation \cite{KoLeOv2}.

Another example of $\Lambda$ that has been overlooked in the
literature \cite{Adam:2011hj, AdQuSaGuWe, BeNeSc, Fr}
is 
\begin{align}
\Lambda = \kappa (\partial_m \Phi \partial^m \Phi \partial_n
 \Phi^{\dagger} \partial^n \Phi^{\dagger})^{-1} 
\frac{1}{(1 + \Phi \Phi^{\dagger})^4} 
\left[
(\partial_m \Phi^{\dagger} \partial^m \Phi)^2 
- 
\partial_m \Phi \partial^m \Phi \partial_n \Phi^{\dagger} \partial^n \Phi^{\dagger}
\right],
\end{align}
where $\kappa$ is a parameter.
Then, with the Fubini-Study metric $K_{\varphi \bar{\varphi}} =
\frac{1}{(1 + |\varphi|^2)^2}$ for the $\mathbb{C}P^1$ model,
the bosonic part of the Lagrangian becomes
\begin{align}
\mathcal{L}_{1b} = - \frac{\partial_m \varphi \partial^m
 \bar{\varphi}}{(1 + |\varphi|^2)^2} 
+ \kappa \frac{(\partial_m \varphi \partial^m \bar{\varphi})^2 -
 |\partial_m \varphi \partial^m \varphi|^2}{(1 + |\varphi|^2)^4}.
\label{eq:FS}
\end{align}
This is nothing but the Faddeev-Skyrme model 
\cite{Faddeev:1996zj}.
The previous trials to construct 
an ${\cal N}=1$ supersymmetric extension 
of the Faddeev-Skyrme model concluded that 
one needs an extra four-derivative term 
containing four time derivatives \cite{BeNeSc, Fr} 
while the Lagrangian in Eq.~\eqref{eq:FS} does not. 
It was discussed in Ref.~\cite{Fr}  that such a term destabilizes Hopfions (knot solitons).
Therefore, the Lagrangian \eqref{eq:Lagrangian} provides an $\mathcal{N} =1$ supersymmetric extension 
of the Faddeev-Skyrme model 
without four time derivatives, which is expected to give 
stable Hopfions.

More generally, since the function $\Lambda$ is completely arbitrary, one can construct 
supersymmetric extension of {\it any} bosonic models that consist of a 
complex scalar field $\varphi$. 
More surprisingly, we further point out 
that it is also possible to introduce an arbitrary scalar potential 
$V(\varphi,\varphi^*)$
even without superpotentials,   
by choosing $\Lambda$ as 
\begin{align}
\Lambda = - (\partial_m \Phi \partial^m \Phi \partial_n
 \Phi^{\dagger} \partial^n \Phi^{\dagger})^{-1} V(\Phi,\Phi^\dagger)  .
\end{align}
However, as we will clarify later, 
superpotentials play an important role when one considers BPS
solutions. 

On the other hand, for the second solution \eqref{eq:second_sol}, 
the bosonic part of the on-shell Lagrangian is  
\begin{align}
\mathcal{L}_{2b} = \left(
\frac{}{}
|\partial_m \varphi \partial^m \varphi |^2 
- (\partial_m \varphi \partial^m \bar{\varphi})^2
\right) \Lambda - \frac{(K_{\varphi \bar{\varphi}})^2}{4 \Lambda}. 
\label{eq:W0_osl2}
\end{align}
In this branch, the canonical kinetic term disappears.\footnote{
When $\Lambda$ is chosen as $\Lambda = - \left(
\frac{}{}
|\partial_m \varphi \partial^m \varphi |^2 
- (\partial_m \varphi \partial^m \bar{\varphi})^2
\right)^{-1} \partial_n \varphi \partial^n \bar{\varphi}$, the canonical
kinetic term recovers. 
However quite non-linear higher derivative terms remain in the
Lagrangian due to the factor $1/\Lambda$.
This possibility was discussed in the context of higher derivative
supergravity models \cite{KoLeOv}.
}
This model was first studied in 
Ref.~\cite{AdQuSaGuWe} where supersymmetric
extensions of the baby Skyrme model are discussed.
We note that the second branch \eqref{eq:W0_osl2} 
does not have the smooth limit to the canonical theory ($\Lambda \to 0$).
Therefore we call this the non-canonical branch.
Since $F \bar{F}$ should be positive semi-definite, 
the second solution \eqref{eq:second_sol} is consistent only in the region 
\begin{align}
- \frac{K_{\varphi \bar{\varphi}}}{2 \Lambda} + \partial_m \varphi
 \partial^m \bar{\varphi} \ge 0.
\label{eq:consistency}
\end{align}

We comment on the last term in Eq.~\eqref{eq:W0_osl2}. 
The term 
$(K_{\varphi \bar{\varphi}})^2/4 \Lambda$ is considered as a scalar potential term since it remains 
when the function $\Lambda$ does not depend on fields with space-time derivatives. 
For a vacuum configuration, the condition \eqref{eq:consistency}
implies $\Lambda < 0$ for the positive definite K\"ahler metric
$K_{\varphi \bar{\varphi}} > 0$. 
Then the scalar potential at a vacuum becomes negative 
even for the manifestly supersymmetric construction of the model.
One resolution of this puzzle is the existence of
ghosts, i.e., fields with a kinetic term of the wrong sign.
However it is not obvious whether ghosts exist or not 
since there is no kinetic term in the Lagrangian \eqref{eq:W0_osl2} and
no consistent free theory is defined.
In that case, $K$ loses its meaning of the K\"ahler potential and
what determines the sign of the potential energy is the function $K$.
When $K_{\varphi \bar{\varphi}}$ is negative, $\Lambda$ and the scalar potential
become positive. 
Actually, choosing the functions of $K$ and $\Lambda$ appropriately, one can
construct scalar potentials that have desired properties
\cite{AdQuSaGuWe}.

\subsubsection{$W \neq 0$ case}

When $W\not=0$ one eliminates $\bar{F}$ in \eqref{eq:aux_eq} and obtains
the equation for the auxiliary field $F$:
\begin{eqnarray}
2 \Lambda (\varphi, \bar{\varphi}) \frac{\partial W}{\partial \varphi} F^3 
+ 
\frac{\partial \bar{W}}{\partial \bar{\varphi}} 
\left(
K_{\varphi \bar{\varphi}} - 2 \Lambda (\varphi, \bar{\varphi}) \partial_m \varphi \partial^m \bar{\varphi}
\right) F
+ 
\left(
\frac{\partial \bar{W}}{\partial \bar{\varphi}}
\right)^2 = 0.
\label{eq:aux_eom}
\end{eqnarray}
When there are no higher derivative corrections $\Lambda = 0$, 
one recovers the ordinary $F$-term solution $F = - \frac{1}{K_{\varphi \bar{\varphi}}}
\frac{\partial \bar{W}}{\partial \bar{\varphi}}$.
Since the equation \eqref{eq:aux_eom} is an algebraic equation of cubic
power, the solutions are obtained by the Cardano's method \cite{SaYaYo},
\begin{eqnarray}
& & F = \omega^k 
\sqrt[3]{
- \frac{q}{2} 
+ \sqrt{\left(\frac{q}{2}\right)^2 + \left(\frac{p}{3}\right)^3}}
+ \omega^{3-k}
\sqrt[3]{
- \frac{q}{2} 
- \sqrt{\left(\frac{q}{2}\right)^2 + \left(\frac{p}{3}\right)^3}}, 
\notag \\
& & k = 0,1,2, \ \omega^3 = 1,
\label{eq:aux_3sol}
\end{eqnarray}
where $\omega$ is a cubic root of unity and $p$ and $q$ are given by 
\begin{eqnarray}
p &=& \frac{1}{2 \Lambda (\varphi, \bar{\varphi})} 
\left(
\frac{\partial W}{\partial \varphi}
\right)^{-1} 
\left(
\frac{\partial \bar{W}}{\partial \bar{\varphi}}
\right)
\left(
K_{\varphi \bar{\varphi}} - 2 \Lambda (\varphi, \bar{\varphi}) \partial_m \varphi \partial^m \bar{\varphi}
\right), \\
q &=& \frac{1}{2 \Lambda (\varphi, \bar{\varphi})} 
\left(
\frac{\partial W}{\partial \varphi}
\right)^{-1} 
\left(
\frac{\partial \bar{W}}{\partial \bar{\varphi}}
\right)^2.
\end{eqnarray}
The on-shell Lagrangian is obtained by substituting the solutions of the
auxiliary field into the Lagrangian \eqref{eq:comLagrangian},
\begin{align}
\mathcal{L}_{b} =& - \frac{\partial^2 K}{\partial \varphi \partial \bar{\varphi}}
 \partial_m \varphi \partial^m \bar{\varphi} 
+ (\partial_m \varphi \partial^m \varphi) (\partial_n \bar{\varphi}
\partial^n \bar{\varphi}) \Lambda 
\notag \\
 & 
+ \tilde{F} \bar{\tilde{F}} (- K_{\varphi \bar{\varphi}}  + 2 \Lambda \partial_m \varphi \partial^m
 \bar{\varphi}) - 3 (\tilde{F} \bar{\tilde{F}})^2 \Lambda,
\label{eq:on-shell_Lagrangian}
\end{align}
where $\tilde{F}$ ($\bar{\tilde{F}}$) is one of the solutions in
\eqref{eq:aux_3sol}.
Therefore, there are three different on-shell branches in this model.
We note that although the model is corrected by higher derivative terms,
the supersymmetry requires correction terms in the scalar potential that
do not contain derivative terms. 
In particular, the scalar potential of the model is 
calculated to be 
\begin{align}
V (\varphi) = |\tilde{F}|^2 (K_{\varphi \bar{\varphi}} + 3 \Lambda^{(0)} |\tilde{F}|^2).
\label{eq:potential}
\end{align}
Here $\Lambda^{(0)}$ is the function $\Lambda$ where $\partial_m \varphi
= 0$. 
We note that even for the manifestly supersymmetric Lagrangian
\eqref{eq:Lagrangian} with the positive K\"ahler metric $K_{\varphi
\bar{\varphi}}$, a negative scalar potential is possible when
$\Lambda^{(0)} < 0$. 
Again, this fact would be an indication of ghost states in the theory.
As we will see below, the on-shell Lagrangian potentially includes ghost states.

Now we examine the structure of the on-shell Lagrangians in each
branch. To see the effects of superpotentials,
we write down the explicit on-shell component Lagrangian. 
In particular, we examine the relation between the positive
definiteness of the scalar potential and the ghost states.
A similar analysis about the scalar potential 
was performed in the context of the
four-dimensional $\mathcal{N} = 1$ supergravity \cite{KoLeOv}, 
in which negative potentials are not problematic.  
On the other hand, 
negative potentials could be problematic 
for the rigid supersymmetric case, 
on which we focus here.

\begin{figure}[t]
\begin{center}
\subfigure[$\Lambda = 1$]
{
\includegraphics[scale=.5]{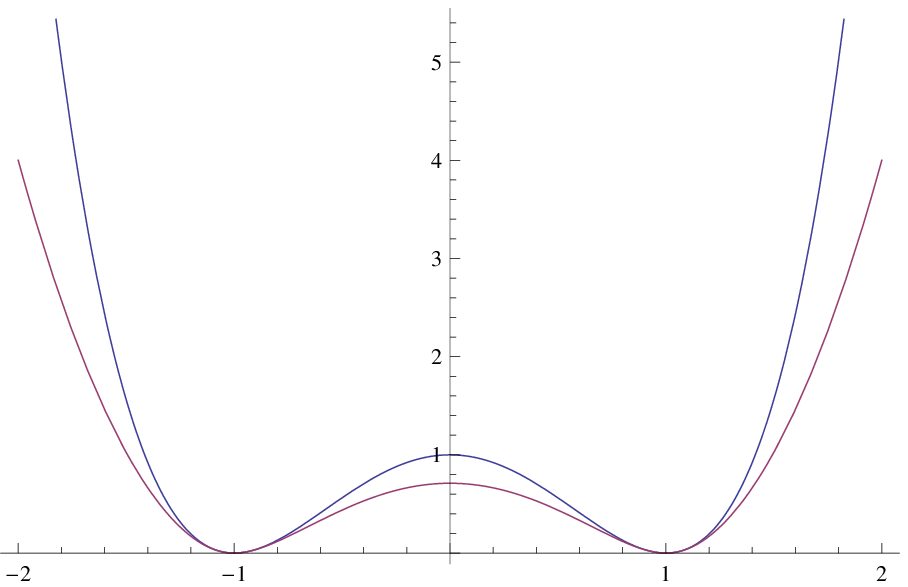}
}
\subfigure[$\Lambda = \varphi \bar{\varphi}$]
{
\includegraphics[scale=.5]{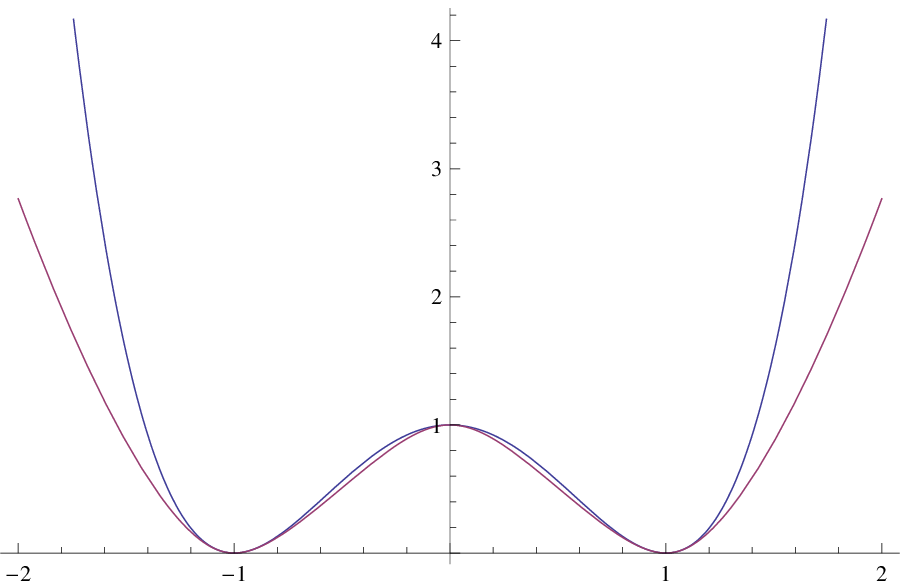}
}
\subfigure[$\Lambda = (\varphi \bar{\varphi})^{-1}$]
{
\includegraphics[scale=.5]{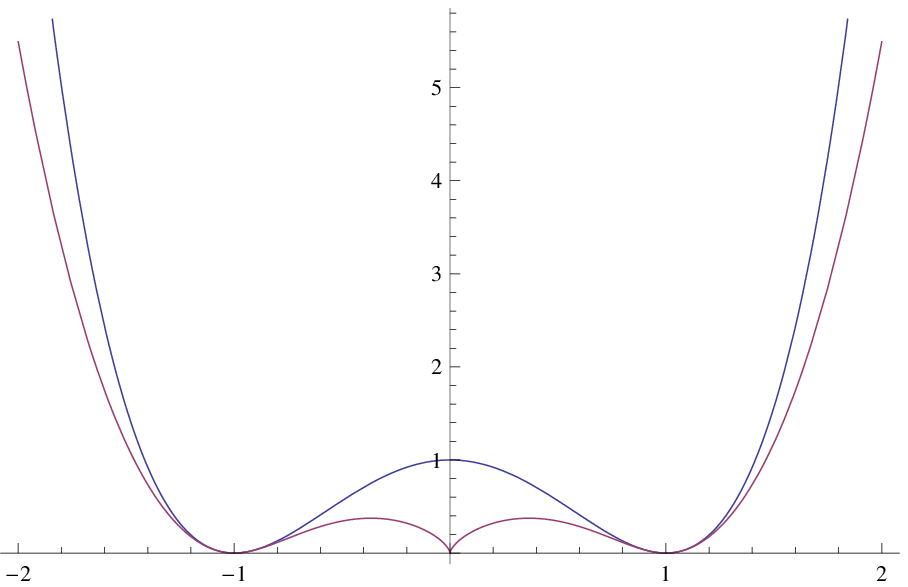}
}
\end{center}
\caption{
Examples of the deformed potentials $V(|\varphi|)$ for $K_{\varphi
 \bar{\varphi}}$, $W = \Phi - \frac{1}{3} \Phi^3$. 
The upper (blue) lines represent the undeformed potentials, 
while the lower (red) lines are deformed ones. The figures correspond to the $k=0$ solution.
}
\label{fig:potential}
\end{figure}

We note that when $W \not= 0$, 
a solution $F=0$ is not allowed. 
We first consider 
the canonical branch 
where the solution of the auxiliary field \eqref{eq:aux_3sol} has
the smooth limit $\Lambda \to 0$ \cite{SaYaYo}.
We look for a perturbative expression of the Lagrangian for small $\Lambda$.
The solution of the auxiliary field is expanded as 
\begin{align}
F = F_0 + \alpha F_1 + \alpha^2 F_2 + \cdots 
\end{align}
where $\alpha$ is a parameter associated with the small $\Lambda$ expansion and
$F_0$ is the solution in $\alpha = 0$ ($\Lambda = 0$). This is given by 
\begin{align}
F_0 = - (K_{\varphi \bar{\varphi}})^{-1} \bar{W}'.
\end{align}
Here $W' = \frac{\partial W}{\partial \varphi}$ and $\bar{W}'$ is the
complex conjugate of $W'$.
The explicit forms of $F_1$ and $F_2$ are obtained iteratively. They are
found to be 
\begin{align}
F_1 =& \ \frac{2 \Lambda \bar{W}'}{(K_{\varphi \bar{\varphi}})^2} 
\left[
\frac{W' \bar{W}'}{(K_{\varphi \bar{\varphi}})^2} 
- \partial_m \varphi \partial^m \bar{\varphi}
\right], \\
F_2 =& \ - \frac{4 \Lambda^2 \bar{W}'}{(K_{\varphi \bar{\varphi}})^7} 
(\bar{W}' W' - K_{\varphi \bar{\varphi}} \partial_m \varphi \partial^m \bar{\varphi}) 
\left\{
3 W' \bar{W}' - (K_{\varphi \bar{\varphi}})^2 \partial_m \varphi
 \partial^m \bar{\varphi}
\right\}.
\end{align}
Then, substituting these solutions into the auxiliary field $F$ in the
Lagrangian \eqref{eq:on-shell_Lagrangian}, we obtain the 
on-shell Lagrangian, (we take $\alpha = 1$ for simplicity)
\begin{align}
\mathcal{L}_b =& - K_{\varphi \bar{\varphi}} \partial_m \varphi \partial^m \bar{\varphi}  
- \frac{2 \Lambda V_0}{K_{\varphi \bar{\varphi}}} \partial_m \varphi \partial^m \bar{\varphi}
+ \frac{8 \Lambda^2 V_0^2}{(K_{\varphi \bar{\varphi}})^3} \partial_m \varphi \partial^m \bar{\varphi} 
\notag \\
& 
+ \Lambda |\partial_m \varphi \partial^m \varphi|^2 
- \frac{4 V_0 \Lambda^2}{(K_{\varphi \bar{\varphi}})^2} (\partial_m \varphi \partial^m \bar{\varphi})^2 
\notag \\
& 
- V_0 
+ \frac{\Lambda V_0^2}{(K_{\varphi \bar{\varphi}})^2}
- \frac{4 \Lambda^2 V_0^3}{(K_{\varphi \bar{\varphi}})^4} 
+ \mathcal{O} (\alpha^4).
\end{align}
Here $V_0 = \frac{1}{K_{\varphi \bar{\varphi}}} | W' |^2$ is the ordinary scalar potential in the
supersymmetric chiral models.
We note that the scalar potential is deformed by the non-zero
$\Lambda$ and the vacuum structure clearly depends on the structure of
the function $\Lambda$.
The examples of the deformed scalar potentials are found in Fig.~\ref{fig:potential}.
The Lagrangian contains an infinite number of the higher
derivative terms that are induced by non-zero $\Lambda$ and $W$. 
The structure of the derivative terms is completely determined by supersymmetry. 
We point out that even for the canonical kinetic term, it is deformed by $\Lambda$.
Up to $\mathcal{O} (\Lambda^2)$, it is given by 
\begin{align}
 \mathcal{L}_K = - 
 \left[
 K_{\varphi \bar{\varphi}} + \frac{2 \Lambda V_0}{K_{\varphi \bar{\varphi}}} 
- \frac{8 \Lambda^2 V_0^2}{(K_{\varphi \bar{\varphi}})^3} 
 \right]
  \partial_m \varphi \partial^m \bar{\varphi}
+ \mathcal{O} (\Lambda^3).
\label{eq:lag-pot}
\end{align}
Since $\Lambda$ is an arbitrary function, the sign of the kinetic
term can be flipped even for the positive  
definite K\"ahler metric
$K_{\varphi \bar{\varphi}}$. If the sign of the kinetic term is changed, there appear ghost
states in the theory \cite{AnDuGh}. 
In that case, the model shows instability caused by the higher
derivatives. This fact leads to the non-positive semi-definite potential
\eqref{eq:potential} even for supersymmetric theories. 
The sign of the kinetic term depends on the explicit forms of the functions $\Lambda$ and $W$.
Although it is important and interesting, 
we do not pursue the (non-)existence of the ghost states in this paper.
We also note that the metric of the 
target space of the nonlinear sigma model 
in the Lagrangian (\ref{eq:lag-pot}) 
does not have to be K\"ahler anymore 
even though it is ${\cal N}=1$ supersymmetric.

Next, we study the effect of the superpotential in the non-canonical branch
associated with the solution \eqref{eq:second_sol}.
Since we cannot take the $\Lambda \to 0$ limit, we consider the small $W$ perturbation
around $W = 0$.
The solution of the auxiliary field is expanded as 
\begin{align}
F = F_0' + \beta F_1' + \beta^2 F_2' + \cdots,
\end{align}
where $\beta$ is a parameter associated with the small $W$ expansion and 
\begin{align}
F_0' = \sqrt{ - \frac{K_{\varphi \bar{\varphi}}}{2 \Lambda} + \partial_m \varphi \partial^m \bar{\varphi}}.
\end{align}
Here we choose a real solution of $F_0$. Using the $U(1)_R$ symmetry, we
make the superpotential be real and positive. 
Then, the solutions $F_1'$ and $F_2'$ are found to be 
\begin{align}
F_1' =& \ - \frac{W'}{4 \Lambda} 
\left(
- \frac{K_{\varphi \bar{\varphi}}}{2 \Lambda} + \partial_m \varphi \partial \bar{\varphi}
\right)^{-1}, \\
F_2' =& \ - \frac{3 (W')^2}{32 \Lambda^2} 
\left(
- \frac{K_{\varphi \bar{\varphi}}}{2 \Lambda} + \partial_m \varphi \partial \bar{\varphi}
\right)^{- \frac{5}{2}}.
\end{align}
The on-shell Lagrangian is 
\begin{align}
\mathcal{L}_b =& \ 
\left(
|\partial_m \varphi \partial^m \bar{\varphi}|^2 - (\partial_m \varphi
 \partial^m \bar{\varphi})^2
\right) \Lambda - \frac{(K_{\varphi \bar{\varphi}})^2}{4 \Lambda} 
\notag \\
& \ - 2 (K_{\varphi \bar{\varphi}} V_0)^{\frac{1}{2}} 
\left(
- \frac{K_{\varphi \bar{\varphi}}}{2 \Lambda} + \partial_m \varphi \partial^m \bar{\varphi}
\right)^{\frac{1}{2}}
- \frac{K_{\varphi \bar{\varphi}} V_0}{16 \Lambda} 
\left(
- \frac{K_{\varphi \bar{\varphi}}}{2 \Lambda} + \partial_m \varphi \partial^m \bar{\varphi}
\right)^{-1}
+ \mathcal{O} (\beta^3).
\end{align}
We can observe that the scalar potential $(K_{\varphi
\bar{\varphi}})^2/4\Lambda$ is deformed by the superpotential $W$.

Finally, a comment is in order.
We started from the four-dimensional theory.
However, the lower dimensional models, such as 
three-dimensional $\mathcal{N} =2$ 
and two-dimensional $\mathcal{N} =(2,2)$ 
theories can be easily obtained by
the dimensional reduction. 
Actually, the $W=0$ case corresponds to
the three-dimensional $\mathcal{N} = 2$ models discussed in Ref.~\cite{AdQuSaGuWe}.

\section{BPS domain walls and their junction}\label{sec:wall}

In this and the next sections, 
we study BPS configurations in the supersymmetric higher
derivative chiral models discussed in the previous section.
Since we consider models with scalar fields, we focus on the BPS domain walls and lumps in the following.

BPS equations that preserve a fraction of supersymmetry are obtained
from the condition that the on-shell supersymmetry transformation of the
fermion vanishes $\delta_{\xi}^{\text{on}} \psi_{\alpha} = 0$.
Here $\delta^{\text{on}}_{\xi}$ represents the on-shell supersymmetry
transformation by parameters $\xi_{\alpha}$, $\bar{\xi}^{\dot{\alpha}}$.
The off-shell supersymmetry transformation $\delta^{\text{off}}_{\xi}$ of the fermion is given by 
\begin{equation}
\delta_{\xi}^{\text{off}} \psi_{\alpha} 
= i \sqrt{2} (\sigma^m)_{\alpha \dot{\alpha}}
\bar{\xi}^{\dot{\alpha}} \partial_m \varphi 
+ \sqrt{2} \xi_{\alpha} F,
\label{eq:SUSY_transformation}
\end{equation}
By substituting a solution of the auxiliary field $F$ into
$\delta^{\text{off}}_{\xi} \psi_{\alpha} = 0$ and assuming  
a specific field configuration together with appropriate Killing
spinor conditions on $\xi_{\alpha}$, $\bar{\xi}^{\dot{\alpha}}$, we find
corresponding on-shell BPS equations. 
Since there is the variety of branches associated with the solutions $F$ in our
model, we study each branch separately.

\subsection{1/2 BPS domain walls}
When a scalar field model with an ordinary canonical kinetic term has a potential with
several 
vacua, there is a domain wall solution that interpolates
between these vacua.
We look for 1/2 BPS domain wall solutions in the higher derivative model \eqref{eq:Lagrangian}.
We consider domain wall configurations of the complex scalar field $\varphi$. 
Namely, the field depends on the one direction $\varphi = \varphi
(x^1)$.
We first consider the case in which the superpotential exists.
In this case, the solution $F=0$ is not allowed. Therefore, we
generically consider the $F \not=0$ branch.
The Killing spinor condition for the 1/2 BPS domain wall configuration is \cite{Dvali:1996xe} 
\begin{eqnarray}
\xi_{\alpha} = - i e^{i \eta} (\sigma^1)_{\alpha \dot{\alpha}}
 \bar{\xi}^{\dot{\alpha}}.
\end{eqnarray}
Here $\eta$ is a phase factor. 
Then the off-shell BPS equation is 
\begin{eqnarray}
\partial_1 \varphi = e^{i \eta } F.
\label{eq:BPS_domain_wall_os}
\end{eqnarray}
By plugging a solution in \eqref{eq:aux_3sol} into the right
hand side of the equation \eqref{eq:BPS_domain_wall_os} and arranging the
resulting condition by $\partial_1 \varphi$, we obtain the on-shell BPS condition.
Here, instead of that, we use the equation of motion for the auxiliary
field $F$ in order to observe the universal property of the three solutions \eqref{eq:aux_3sol}. 
Substituting the BPS condition \eqref{eq:BPS_domain_wall_os} into the equation of motion for $F$, we obtain
\begin{align}
K_{\varphi \bar{\varphi}} e^{-i\eta} \partial_1 \varphi + 
\left\{
- 2 e^{-i \eta} \partial_1 \varphi \cdot \partial_1 \varphi \partial_1
 \bar{\varphi}
+ 2 e^{-2i\eta} (\partial_1 \varphi)^2 e^{i\eta} \partial_1 \bar{\varphi}
\right\} 
\Lambda 
+ \frac{\partial \bar{W}}{\partial \bar{\varphi}} 
= 0.
\end{align}
The higher derivative terms including $\Lambda$ cancel out and we obtain
the on-shell BPS equation
\begin{align}
K_{\varphi \bar{\varphi}} \partial_1 \varphi + e^{i\eta} \frac{\partial \bar{W}}{\partial
 \bar{\varphi}} = 0.
\label{eq:BPS_domain_wall}
\end{align}
Equation.~\eqref{eq:BPS_domain_wall} is 
nothing but the ordinary (without higher derivative terms) BPS condition for the domain wall.
This result suggests that 
even for the existence of the three different on-shell branches in the model, the BPS
domain wall cannot distinguish them. 
Furthermore, 
the on-shell energy density of the domain wall is evaluated as 
\begin{align}
\mathcal{E} =& \ K_{\varphi \bar{\varphi}} |\partial_1 \varphi |^2 - |\partial_1 \varphi |^4 \Lambda
- | \partial_1 \varphi |^2 (- K_{\varphi \bar{\varphi}} + 2 \Lambda |\partial_1 \varphi |^2) +
 3 |\partial_1 \varphi |^4 \Lambda 
\notag \\
=& - e^{-i\eta} \partial_1 W + h.c.
\end{align}
The last expression gives the tension of the ordinary BPS domain wall.
Therefore we conclude that all the higher derivative corrections 
to the solutions and energy are canceled out in the BPS domain walls.
This is a consequence of the fact that the configuration depends on the one direction.
It is easy to confirm that the solutions to the BPS condition 
\eqref{eq:BPS_domain_wall_os} together with the equation of motion for
the auxiliary field \eqref{eq:BPS_domain_wall} satisfy the full equation
of motion for the scalar field
\footnote{We have assumed that $\Lambda$ does not depends on the
second space-time derivatives or higher of $\varphi$.} 
\begin{align}
& - \frac{\partial^3 K}{\partial \varphi \partial^2 \bar{\varphi}}
 ( |\partial_m \varphi|^2 - |F|^2) + \frac{\partial^2 \bar{W}}{\partial
 \bar{\varphi}^2} \bar{F} 
+ \left[
|\partial_m \varphi \partial^m \varphi |^2 - 2 |F|^2 |\partial_m
 \varphi|^2 + |F|^4
\right] \frac{\partial \Lambda}{\partial \bar{\varphi}} 
\notag \\
& - \partial_m 
\left[
- K_{\varphi \bar{\varphi}}
 \partial^m \varphi
+ 2 \Lambda ( (\partial_n \varphi)^2 \partial^m \bar{\varphi} - |F|^2
 \partial^m \varphi)
+ 
\left\{
|\partial_n \varphi \partial^n \varphi|^2 - 2 |F|^2 |\partial_n
 \varphi|^2 + |F|^4
\right\} \frac{\partial \Lambda}{\partial (\partial_m \bar{\varphi})}
\right] = 0.
\label{eq:varphi_eom}
\end{align}

Next, we consider the case in which $W = 0$. Even for this case, 
there is the scalar potential $(K_{\varphi \bar{\varphi}})^2/4 \Lambda$
in the Lagrangian \eqref{eq:W0_osl2}. This branch corresponds to the
$F\not=0$ solution \eqref{eq:second_sol}. 
Substituting the off-shell BPS condition \eqref{eq:BPS_domain_wall_os} into the equation of motion for
the auxiliary field \eqref{eq:eom_auxiliary_no_supot} and assuming
$F\not=0$, the on-shell BPS condition becomes
\begin{align}
K_{\varphi \bar{\varphi}} = 0.
\label{eq:NC_dw}
\end{align}
This condition never provides the domain wall equation.
When there is no ghost, Eq.~\eqref{eq:NC_dw} 
is just a vacuum condition of the scalar potential
 $(K_{\varphi \bar{\varphi}})^2/4 \Lambda$.
Therefore, 
although there is a scalar
potential in the non-canonical branch, superpotentials are necessary for
1/2 BPS domain wall solutions.
We also note that the 1/2 BPS domain wall solution to the equation
\eqref{eq:BPS_domain_wall} interpolates between ``vacua'' specified by the
superpotential $W' = 0$ as its tension stands for.
We stress that the condition $W' = 0$ does not always imply vacua of the
scalar potential especially in the non-canonical branch.
The BPS domain walls remain intact even for the deformation of the scalar potential.
Although there are other vacua that originate from the singularity of
the function $\Lambda$ (see Fig.1 (c)), domain walls that interpolate
these vacua are not BPS and break all the supersymmetry.

\subsection{1/4 BPS domain wall junctions}
We next consider 1/4 BPS domain wall junctions 
\cite{GiTo}.
The scalar field depends on the two spacial directions $x^1 and x^2$.
First, we consider the $W \not= 0$ case. 
We impose the Killing spinor conditions on the supersymmetry parameters,
\begin{align}
\frac{1}{2} (\sigma^1 + i \sigma^2)_{\alpha \dot{\alpha}} \bar{\xi}^{\dot{\alpha}} =
 0, \qquad 
\frac{1}{2} (\sigma^1 - i \sigma^2)_{\alpha \dot{\alpha}}
 \bar{\xi}^{\dot{\alpha}} = i e^{- i \eta} \xi_{\alpha},
\label{eq:Killing_spinor}
\end{align}
where $\eta$ is a phase factor.
Then we obtain the BPS condition from the supersymmetry transformation
\eqref{eq:SUSY_transformation},
\footnote{
We define the complex coordinate $z = \frac{1}{2} (x^1 + i x^2)$ and
derivatives $\partial = \frac{\partial}{\partial z} = \partial_1 - i
\partial_2$. $\bar{\partial}$ is the complex conjugate of $\partial$.
}
\begin{align}
\bar{\partial} \varphi = e^{i\eta} F.
\label{eq:BPS_junction}
\end{align}
Here $F$ in the right hand side is one of the solutions in \eqref{eq:aux_3sol}.
This is the 1/4 BPS condition. 
Substituting the condition \eqref{eq:BPS_junction} into the equation of
motion \eqref{eq:aux_eq} for the auxiliary field, we obtain the on-shell BPS equation on
the scalar field:
\begin{align}
K_{\varphi \bar{\varphi}} \bar{\partial} \varphi - \bar{\partial} \varphi 
(|\partial \varphi|^2 - |\bar{\partial} \varphi|^2) \Lambda +
 e^{i \eta} \frac{\partial \bar{W}}{\partial \bar{\varphi}} = 0.
\label{eq:on-shell_BPS_junction}
\end{align}
When $\Lambda = 0$, the on-shell BPS equation
\eqref{eq:on-shell_BPS_junction} becomes that of the
ordinary BPS domain wall junctions \cite{GiTo} of which the analytic
solutions are studied in Ref.~\cite{NaNiSa}.
Different from the 1/2 BPS domain wall case, the higher derivative corrections
do not cancel in the equation \eqref{eq:on-shell_BPS_junction}.
The solutions are deformed from the ones in Ref.~\cite{NaNiSa} in general
and depend on the explicit form of the function $\Lambda$.

Now we confirm that the BPS solutions to \eqref{eq:BPS_junction} satisfy
the full equation of motion for the scalar field \eqref{eq:varphi_eom}.
Using the BPS condition \eqref{eq:BPS_junction}, we find the following
terms in \eqref{eq:varphi_eom} vanish:
\begin{align}
|\partial_m \varphi \partial^m \varphi |^2 - 2 |F|^2 |\partial_m
 \varphi|^2 + |F|^4 = 0.
\end{align}
By using the BPS equation and the equation of motion for the auxiliary
field, we find that the other terms in \eqref{eq:varphi_eom} also vanish:
\begin{align}
& - \frac{\partial^3 K}{\partial \varphi \partial^2 \bar{\varphi}}
 ( |\partial_m \varphi|^2 - |F|^2) + \frac{\partial^2 \bar{W}}{\partial
 \bar{\varphi}^2} \bar{F} 
\notag \\
& - \partial_m 
\left[
- \frac{\partial^2 K}{\partial \varphi \partial \bar{\varphi}}
 \partial^m \varphi
+ 2 \Lambda ( (\partial_n \varphi)^2 \partial^m \bar{\varphi} - |F|^2
 \partial^m \varphi)
\right] = 0.
\end{align}
Therefore we conclude that the solutions to the deformed BPS equation
\eqref{eq:on-shell_BPS_junction} actually satisfy the full equation of
motion  in Eq.~\eqref{eq:varphi_eom}.

The energy density of the domain wall junction is evaluated as 
\begin{align}
\mathcal{E} =& \ K_{\varphi \bar{\varphi}} \partial_i \varphi \partial_i \bar{\varphi} - 
(\partial_i \varphi \partial_i \varphi) (\partial_j \bar{\varphi}
 \partial_j \bar{\varphi}) \Lambda 
- |F|^2 (-K_{\varphi \bar{\varphi}} + 2 \Lambda \partial_i \varphi \partial_i
 \bar{\varphi}) + 3 |F|^4 \Lambda 
\notag \\
=& \ 
\frac{1}{2} K_{\varphi \bar{\varphi}} (|\partial \varphi|^2 - |\bar{\partial} \varphi|^2) - 2
 \mathrm{Re} 
\left[
e^{- i \eta} \frac{\partial W}{\partial \bar{z}}
\right].
\end{align}
This is nothing but the expression of the ordinary (without higher
derivative terms) domain wall junctions.
After integration over the $(x^1,x^2)$ plane, the first term gives the
junction charge and the second term gives the tension of the domain walls.
They are evaluated on the boundary at the infinity of the $(x^1, x^2)$
plane.
Again, the junction charge and the domain wall tension are solely determined by
the asymptotic boundary conditions of the scalar field and the superpotential,
and do not depend on the function $\Lambda$.
Although the expression of the Bogomol'nyi bound of the energy is not
deformed by the higher derivative terms, we stress that the solutions of the 1/4 BPS
domain wall junction are potentially deformed in general.

Finally we examine 1/4 BPS domain wall junctions in the $W = 0$ non-canonical branch. 
The Killing spinor and the off-shell BPS conditions are 
given by Eqs.~\eqref{eq:Killing_spinor} and \eqref{eq:BPS_junction}.
The solution of the auxiliary field is given in Eq.~\eqref{eq:second_sol}.
Then, the on-shell BPS equation is found to be 
\begin{align}
\frac{1}{2} (|\partial \varphi|^2 - |\bar{\partial} \varphi|^2) =
 \frac{K_{\varphi \bar{\varphi}}}{2 \Lambda}.
\label{eq:non-canonical_dw_junction}
\end{align}
Eq.~\eqref{eq:non-canonical_dw_junction} is supplemented by the
consistency condition in Eq.~\eqref{eq:consistency}.
Again, the higher derivative corrections to the
on-shell 1/4 BPS condition are not canceled.
We will comment on this equation in the next section.

\section{BPS lumps and baby Skyrmions} \label{sec:lump}

Next we consider lumps in $W= 0$ higher derivative models.
We look for the BPS equation for lumps that depend on $x^1$ and $x^2$. 
Recall that for the $W = 0$ case, the solutions of the auxiliary field
are given by 
\begin{align}
F =& \ 0, 
\label{eq:sol1}
\\
F  =& \ 
e^{i \alpha}
\sqrt{
- \frac{K_{\varphi \bar{\varphi}}}{2 \Lambda} + \partial_m \varphi \partial^m \bar{\varphi}
},
\label{eq:sol2}
\end{align}
where $\alpha$ is a phase factor. 
There are the canonical and non-canonical branches associated with the
solutions in Eqs.~\eqref{eq:sol1} and \eqref{eq:sol2},  respectively. 
In the following subsections, we examine BPS lump equations in each branch.

\subsection{1/2 BPS lumps}
We first focus on the canonical branch associated with the solution \eqref{eq:sol1}. 
Hopfions in the supersymmetric higher derivative 
$\mathbb{C}P^1$ model of this type 
were discussed before \cite{BeNeSc, Fr}.
BPS lumps in the supersymmetric higher derivative 
$\mathbb{C}P^1$ model  
were discussed in Ref.~\cite{Eto:2012qda}.  
BPS lumps in the higher derivative $\mathbb{C}P^n$ non-linear sigma models were discussed in a different
context without supersymmetry 
\cite{Ferreira:2008nn}. 
 
In this branch, the BPS lump equation is obtained by imposing 
 the first condition in \eqref{eq:Killing_spinor} on the spinor
$\bar{\xi}^{\dot{\alpha}}$, 
as can be seen in, e.g., Refs.~\cite{Shifman:2007ce,Eto:2006pg,Eto:2005sw}. 
Then, the BPS equation for lumps is given by \cite{Polyakov:1975yp}
\begin{align}
\bar{\partial} \varphi =& 0.
\label{eq:lump1}
\end{align}
This is nothing but the ordinary 1/2 BPS lump condition.
This is confirmed by the Bogomol'nyi bound of the energy density.
For the canonical branch, we have 
the energy density
\begin{align}
\mathcal{E} =& \ K_{\varphi \bar{\varphi}} |\partial_i \varphi|^2 - |\partial_i \varphi
 \partial_i \varphi|^2 \Lambda 
\notag \\
=& \ 
|\bar{\partial} \varphi|^2 
\left(
K_{\varphi \bar{\varphi}} - |\partial \varphi|^2 \Lambda
\right) - i K_{\varphi \bar{\varphi}} \varepsilon_{ij} \partial_i
 \varphi \partial_j \bar{\varphi}
\notag \\
\ge& \ - i K_{\varphi \bar{\varphi}} \varepsilon_{ij} \partial_i \varphi
 \partial_j \bar{\varphi}, 
\label{eq:BPS-bound-lump}
\end{align}
where we have assumed the condition 
$\Lambda \le K_{\varphi \bar{\varphi}}/|\partial \varphi|^2$ 
for the positive-semi definiteness of the energy $\mathcal{E}$.
The right hand side is nothing but the topological charge density for
the 1/2 BPS lump. The energy bound is saturated provided the condition
\eqref{eq:lump1} is satisfied.
Then we find that the higher derivative corrections to the solutions and
the energy bound are canceled out in this branch.
It is confirmed that solutions to the equation \eqref{eq:lump1} satisfy
the full equation of motion for the scalar field \eqref{eq:varphi_eom}.
When we consider the Fubini-Study metric for the ${\mathbb C}P^1$ model
and take the function $\Lambda$ as 
\begin{align}
K_{\varphi \bar{\varphi}} = \frac{1}{(1 + |\varphi|^2)^2}, \qquad 
\Lambda= \frac{1}{(1+|\varphi|^2)^4}
\label{eq:CP1}
\end{align}
the bound \eqref{eq:BPS-bound-lump} 
becomes just the BPS bound obtained 
in the context of the effective theory on a 
non-Abelian vortex \cite{Eto:2012qda}. 

In summary, although the Lagrangian contains higher derivative corrections, 
the 1/2 BPS lump solution to the equation \eqref{eq:lump1} (which is a
holomorphic function with appropriate boundary conditions) does not
receive any corrections in the canonical branch \eqref{eq:W0_osl1}.


\subsection{1/4 BPS lumps as compact baby Skyrmions}
We next consider the non-canonical branch.
Since this is associated with the $F\not=0$ solution \eqref{eq:sol2}
even for $W=0$, we need to impose both of the two conditions in
\eqref{eq:Killing_spinor} in order to obtain the BPS equation from the variation of the fermion.
Then the BPS equation is 
\begin{align}
\bar{\partial} \varphi =& e^{i \eta'} 
\sqrt{- \frac{K_{\varphi \bar{\varphi}}}{2 \Lambda} + \frac{1}{2} (|\partial \varphi|^2 +
 |\bar{\partial}\varphi|^2)
}, 
\label{eq:lump2}
\end{align}
where $\eta' = \eta + \alpha \in \mathbb{R}$ is a phase factor.
This is the 1/4 BPS equation.
Again, the BPS lump does not cancel the higher derivative corrections
generally.
We can make the deformed BPS equation \eqref{eq:lump2} into the
following form,
\begin{align}
\frac{1}{2} (|\partial \varphi|^2 - |\bar{\partial} \varphi|^2) =
 \frac{K_{\varphi \bar{\varphi}}}{2 \Lambda}.
\label{eq:lump2-2}
\end{align}
 
We confirm that the solutions to the BPS equation \eqref{eq:lump2} 
satisfy the full on-shell equation of motion for the scalar field
\eqref{eq:varphi_eom}. 
In the non-canonical branch, we have the Bogomol'nyi completion of the
energy 
\begin{align}
\mathcal{E} =& \ 
- \left(
|\partial_i \varphi \partial_i \varphi|^2 - (\partial_i \varphi
 \partial_i \bar{\varphi})^2
\right) \Lambda + \frac{(K_{\varphi \bar{\varphi}})^2}{4 \Lambda} 
\notag \\
=& \ 
\Lambda 
\left[
\frac{1}{2} 
(|\partial \varphi|^2 - |\bar{\partial} \varphi|^2) 
- \frac{K_{\varphi \bar{\varphi}}}{2 \Lambda}
\right]^2 + \frac{K_{\varphi \bar{\varphi}}}{2} (|\partial \varphi|^2 - |\bar{\partial}
 \varphi|^2)
\notag \\
\ge & \ - i K_{\varphi \bar{\varphi}} \varepsilon_{ij} \partial_i
 \varphi \partial_j \bar{\varphi}.
\label{eq:lump_energy_bound}
\end{align}
Since we have $\Lambda > 0$ 
for static configurations 
from the consistency condition
\eqref{eq:consistency} of the solution, 
the energy bound is saturated by the topological
charge density of lumps provided that the BPS condition 
\eqref{eq:lump2-2} is satisfied.
It is obvious that the expression of the topological charge is not corrected by the
higher derivative terms.

In the non-canonical branch, 
the Lagrangian does not
contain ordinary canonical kinetic term.
An example of such a kind of non-canonical model is the extremal (BPS) baby Skyrme model.
The model consists of the fourth derivative term and potential terms in
(2+1) dimensions. More concretely, if we take the K\"ahler potential and
$\Lambda$ as in \eqref{eq:CP1},
then the Lagrangian \eqref{eq:W0_osl2} is nothing but the fourth
derivative part of the baby Skyrme model with an irrelevant constant term.
In Ref.~\cite{AdQuSaGuWe} the
authors found specific K\"ahler potentials and constructed the potentials of the
baby Skyrme model. 
Actually, the condition \eqref{eq:lump2-2} was  
first found in the
supersymmetric baby Skyrme model \cite{AdQuSaGuWe}.
Eq.~\eqref{eq:lump2-2} is the same as the one found in the previous 
section, Eq.~\eqref{eq:non-canonical_dw_junction}.
The difference of solutions is specified by boundary conditions.
However, the energy bound in \eqref{eq:lump_energy_bound} suggests that
there are no BPS domain walls junctions in the non-canonical branch. 
The example of solutions to Eq.~\eqref{eq:lump2-2} are 
the compact baby Skyrmions 
 \cite{Adam:2009px,AdRoSaGuWe} that are solitons
with compact support. 
The BPS states in the higher derivative chiral models are summarized in
Table \ref{tb:BPS_states}.
\begin{table}[t]
\begin{center}
\begin{tabular}{|l||c|c|} 
\hline
 & 1/2 BPS & 1/4 BPS \\ 
\hline \hline
$W=0$ & Lumps ($F=0$) & Compact lumps ($F\not=0$)\\ 
\hline
$W\not=0$ & Domain walls ($F\not=0$) & Deformed domain wall junctions ($F\not=0$)
		 \\ 
\hline
\end{tabular}
\caption{BPS states in the $W=0$
 and $W\not=0$ higher derivative chiral models.
Corresponding solutions of the auxiliary field $F$ (whether they vanish or not)
 are also presented.}
\end{center}
\label{tb:BPS_states}
\end{table}


\section{Conclusion and discussions} \label{sec:conc}
In this paper we have studied BPS states in the four-dimensional
$\mathcal{N} = 1$ supersymmetric higher 
derivative chiral model of which the Lagrangian is given in Eq.~\eqref{eq:Lagrangian}.
The model is governed by 
a (2,2) K\"ahler tensor $\Lambda_{ij\bar k \bar l}$ 
symmetric in holomorphic and anti-holomorphic indices, 
in addition to the K\"ahler potential $K$ and the
superpotential $W$. They are functions of the chiral superfields
$\Phi^i$. 
In particular, the tensor $\Lambda_{ij\bar k \bar l}$ determines the
higher derivative interactions of the models. 
A specific feature of the model is that the auxiliary fields $F^i$ 
do not have space-time derivatives on them and can be eliminated by their equation of motion 
algebraically. 
One can explicitly write down the on-shell Lagrangian of the model at least for a single chiral superfield. 
Since the equation of motion for the auxiliary fields is no longer 
a linear equation, there are several on-shell branches in this model.
This fact deserves new non-trivial BPS equations that include higher
derivative corrections.

When there is no superpotential, there are two distinct on-shell branches. 
One is the canonical branch associated with the
solution $F=0$. An example of this model is the supersymmetric DBI model
\cite{RoTs}. 
We have shown that this branch, in fact, allows a supersymmetric extension
of any bosonic models of complex scalar fields.
We have exhibited the explicit function $\Lambda$ which corresponds to
the supersymmetric extension of the Faddeev-Skyrme model without four time derivatives, 
which is in contrast to the previous studies 
\cite{BeNeSc,Fr}
concluding that such a term is necessary for 
supersymmetry. 
The other branch is the non-canonical one corresponding to
the solution $F\not=0$. In this branch, the ordinary canonical kinetic
term disappears and the Lagrangian starts from the forth order
derivative terms. An example of this model is the extremal (BPS) baby Skyrme model. 
This branch was discussed in Refs.~\cite{AdQuSaGuWe, KhLeOv}.
Although the $W=0$ case has been essentially discussed in the literature
\cite{AdQuSaGuWe, KhLeOv}, 
things get more involved when one introduces a superpotential $W$. 
In this case, a solution $F=0$ is not allowed.
There are three on-shell branches associated with the three different
solutions of the auxiliary field equation \cite{SaYaYo,KoLeOv, FaKe}. 
The resulting on-shell Lagrangians have highly non-linear expressions.
Perturbative analysis reveals the possibility of ghost kinetic terms and
deformations of the scalar potential.

Even though the on-shell Lagrangian is complicated and becomes highly
non-linear 
in the 
$W \not= 0$ case, one can derive the off-shell BPS
conditions from the supersymmetry transformation of fermions.
These conditions are supplemented by the equation of motion for the
auxiliary field giving rise to the on-shell conditions.
We have analyzed the properties of BPS states.
For the 1/2 BPS domain wall case, the higher derivative corrections are
exactly canceled out in the $W\not=0$ case.
The solution to the BPS equation satisfies the full equation of motion
for the scalar field.
We have shown that the tension of the domain wall does not receive any
higher derivative corrections.
In the $W=0$ non-canonical branch, the 1/2 BPS condition does not
provide the domain wall equation.
For the 1/4 BPS domain wall junction in the $W\not=0$ case, the on-shell BPS equation
receives higher derivative corrections.
This is a new 1/4 BPS equation for domain wall junctions.
The solution is deformed by the higher derivative effects and it is
confirmed that the solution satisfies the full equation of motion. 
The expression of the energy bound is shown to be the same as the
ordinary (without higher derivative terms) theory, namely, the sum of
the junction charge and the tension.
For lump configurations in the $W=0$ case, there are two on-shell BPS equations.
One is the 1/2 BPS lumps associated with the $F = 0$ solution, where all the
derivative corrections are canceled out.
The other is the 1/4 BPS lumps associated with the $F\not=0$ solution.
The on-shell BPS equation is deformed by the higher derivative corrections.
This is nothing but the equation studied in Ref.~\cite{AdQuSaGuWe}. 
An example of solutions to this equation is compactons in the extremal
(BPS) baby Skyrme model.

While we were able to solve explicitly 
auxiliary field equations 
(\ref{eq:af-eom})
 only for one chiral superfield, 
reducing the third order algebraic equation 
(\ref{eq:af-eom2}),
the multicomponent equation
(\ref{eq:af-eom}) has yet to be solved.
When the target space has a large isometry, 
it should be possible to solve it.
Construction of more general target spaces, 
for instance a higher derivative ${\mathbb C}P^n$ model 
and its BPS solitons remains as a future problem.

While we have exhausted all BPS states 
that are already 
known in conventional ${\cal N}=1$ supersymmetric theories 
without higher derivatives, 
there may still remain   
unknown BPS states particular for 
higher derivative theories. 
In fact, 1/4 baby BPS Skyrmions do not exist in 
conventional theories. 
A sine-Gordon kink inside 
a domain wall (corresponding to 
a baby Skyrmion in the bulk) \cite{Nitta:2012xq},
a baby Skyrmion inside a domain wall 
(corresponding to a three-dimensional Skyrmion in the bulk)
\cite{Nitta:2012wi}, or  
a baby Skyrmion string 
ending on a domain wall \cite{Gauntlett:2000de} 
or 
stretched between 
domain walls \cite{Isozumi:2004vg,Eto:2006pg}
is one of possibilities 
of composite BPS states. 

It should be important to generalize 
our formalism to theories with extended 
supersymmetries such as
eight supercharges. 
Although, only four out of eight supercharges are manifestly realized in
the $\mathcal{N} = 1$ superfield formalism, this is still useful to study the
off-shell effective theory of BPS solitons in models with eight
supercharges \cite{Eto:2006uw}.
Supersymmetric theories with eight supercharges 
are known to admit plenty of composite 
BPS states \cite{Eto:2006pg,Shifman:2007ce}.
In particular,  
a classification of all possible BPS states 
in supersymmetric theories with eight supercharges 
was given in Ref.~\cite{Eto:2005sw}.
It is an interesting future problem to explore 
which BPS states (do not) receive 
higher derivative corrections.

As this problem concerns, 
1/2 BPS topological solitons 
in theories with eight supercharges 
preserve four supercharges 
on their world-volume.
Off-shell effective actions 
of the 1/2 BPS domain wall and vortex 
were obtained in $d=3+1$, ${\cal N}=1$ superfield formalism 
at the leading order \cite{Eto:2006uw}. 
The formulation presented in this paper 
should be useful to obtain 
the off-shell action of higher derivative corrections 
to these effective actions. 
For instance, 
as mentioned in Eqs.~(\ref{eq:lump1}) and (\ref{eq:BPS-bound-lump}), 
the ${\mathbb C}P^1$ model with 
a four-derivative term appearing as 
the effective theory of a non-Abelian vortex 
admits 1/2 BPS lumps \cite{Eto:2012qda}, 
corresponding to Yang-Mills instantons in the bulk 
\cite{Eto:2004rz}. 
In the same way, 
an $SU(N)$ principal chiral model with 
the Skyrme term appears 
\cite{Eto:2005cc} 
on a non-Abelian domain wall 
\cite{Shifman:2003uh}.
The off-shell higher derivative corrections 
to the effective theories on these solitons 
are some of future directions.

\subsection*{Acknowledgments}
The authors would like to thank Masahide Yamaguchi for discussions.
The work of M.\ N.\ is supported in part by Grant-in-Aid for 
Scientific Research (No. 25400268) and by the ``Topological 
Quantum Phenomena''  Grant-in-Aid for Scientific Research on 
Innovative Areas (No. 25103720) from the Ministry of Education, 
Culture, Sports, Science and Technology  (MEXT) of Japan.
The work of S.~S. is supported in part by Kitasato University Research Grant for Young
Researchers.

\begin{appendix}
\section{Notation and conventions}\label{sec:notation}

We use the notation of the textbook of 
Wess and Bagger \cite{Wess:1992cp}. 
The component expansion of the $\mathcal{N} = 1$ chiral superfield in
the $x$-basis is 
\begin{equation}
\Phi (x, \theta, \bar{\theta}) = \varphi 
+ i \theta \sigma^m \bar{\theta} \partial_m \varphi + \frac{1}{4}
\theta^2 \bar{\theta}^2 \Box \varphi + \theta^2 F,
\end{equation}
where 
only the bosonic components are presented.
The supercovariant derivatives are defined as 
\begin{eqnarray}
D_{\alpha} = \frac{\partial}{\partial \theta^{\alpha}} + i
 (\sigma^m)_{\alpha \dot{\alpha}} \bar{\theta}^{\dot{\alpha}}
 \partial_m, \quad 
\bar{D}_{\dot{\alpha}} = - \frac{\partial}{\partial
\bar{\theta}^{\dot{\alpha}}} - i \theta^{\alpha} (\sigma^m)_{\alpha
\dot{\alpha}} \partial_m.
\end{eqnarray}
The sigma matrices are $\sigma^m = (\mathbf{1}, \vec{\tau})$.
Here $\vec{\tau} = (\tau^1, \tau^-2, \tau^3)$ are Pauli matrices.

The bosonic component of the supercovariant derivatives of $\Phi^i$ are 
\begin{align}
D^{\alpha} \Phi^i D_{\alpha} \Phi^j =& \ 
- 4 \bar{\theta}^2 \partial_m \varphi^i \partial^m \varphi^j 
+ 4 i (\theta \sigma^m \bar{\theta}) (\partial_m \varphi^i F^j + F^i
 \partial_m \varphi^j) 
- 4 \theta^2 F^i F^j 
\notag \\
& \ + 2 \theta^2 \bar{\theta}^2 
\left(
\Box \varphi^i F^j + F^i \Box \varphi^j - \partial_m \varphi^i
 \partial^m F^j - \partial_m F^i \partial^m \varphi^j
\right), \\
\bar{D}_{\dot{\alpha}} \Phi^{\dagger\bar{i}} \bar{D}^{\dot{\alpha}}
 \Phi^{\dagger\bar{j}} =& \ 
- 4 \theta^2 \partial_m \bar{\varphi}^{\bar{i}} \partial^m
 \bar{\varphi}^{\bar{j}} 
- 4 i (\theta \sigma^m \bar{\theta}) (\partial_m \bar{\varphi}^{\bar{i}}
 \bar{F}^{\bar{j}} + \bar{F}^{\bar{i}} \partial_m
 \bar{\varphi}^{\bar{j}}) 
+ 4 \bar{\theta}^2 \bar{F}^{\bar{i}} \bar{F}^{\bar{j}}
\notag \\
& \ 
+ 2 \theta^2 \bar{\theta}^2 
\left(
\bar{F}^{\bar{i}} \Box \bar{\varphi}^{\bar{j}} + \Box
 \bar{\varphi}^{\bar{i}} \bar{F}^{\bar{j}} 
- \partial_m \bar{\varphi}^{\bar{i}} \partial^m \bar{F}^{\bar{j}}
- \partial_m \bar{F}^{\bar{i}} \partial^m \bar{\varphi}^{\bar{j}}
\right), 
\\
D^{\alpha} \Phi^i D_{\alpha} \Phi^k \bar{D}_{\dot{\alpha}}
 \Phi^{\dagger\bar{j}} \bar{D}^{\dot{\alpha}}
 \Phi^{\dagger\bar{l}} 
=& \ 16 \theta^2 \bar{\theta}^2 
\left[
\frac{}{}
(\partial_m \varphi^i \partial^m \varphi^k) (\partial_m
 \bar{\varphi}^{\bar{j}} \partial^m \bar{\varphi}^{\bar{l}})
\right. 
\notag \\
& 
\left.
- \frac{1}{2} 
\left(
\partial_m \varphi^i F^k + F^i \partial_m \varphi^k 
\right)
\left(
\partial^n \bar{\varphi}^{\bar{j}} \bar{F}^{\bar{l}} 
+ \bar{F}^{\bar{j}} \partial^n \bar{\varphi}^{\bar{l}}
\right)
+ F^i \bar{F}^{\bar{j}} F^k \bar{F}^{\bar{l}}
\right].
\end{align}

\end{appendix}


\end{document}